\documentclass[amsmath,amssymb,prl,hyperlink,twocolumn,showpacs,keywords]{revtex4}

	\usepackage{graphicx}
	\usepackage{soul}
	\usepackage[colorlinks=true,citecolor=blue,linkcolor=blue]{hyperref}
	\usepackage[usenames]{color}
	\usepackage{amsfonts}
	\usepackage{color}
	\usepackage{booktabs}
	\usepackage{multirow}
	\usepackage{float}

\begin{document}

	\title{Generation of nondegenerate narrow-band photon pairs for hybrid quantum network }
	
	\author{Jian Wang, Peng-YinJie Lv, Jin-Ming Cui, Bi-Heng Liu, Jian-Shun Tang} \author{Yun-Feng Huang}
\email{hyf@ustc.edu.cn}
\author{Chuan-Feng Li}
\email{cfli@ustc.edu.cn}
\author{Guang-Can Guo}
	
\affiliation{Key Laboratory of Quantum Information, University of Science and Technology of China, CAS, Hefei, 230026, People's Republic of China \\Synergetic Innovation Centre in Quantum Information and Quantum Physics, University of Science and Technology of China, Hefei, Anhui 230026, China}

	
	\begin{abstract}
In a hybrid quantum network, linking two kinds of quantum nodes through photonic channels requires excellent matching of central frequency and bandwidth between both nodes and their interfacing photons. However, pre-existing photon sources can not fulfill this requirement. Using a novel conjoined double-cavity strategy, we report the generation of nondegenerate narrow-band photon pairs by cavity-enhanced spontaneous parametric down-conversion. The central frequencies and bandwidths of the signal and idler photons are independently set to match with trapped ions and solid-state quantum memories. With this source we achieve the bandwidths and central frequencies of 4 MHz at 935 nm and 5 MHz at 880 nm for the signal and idler photons respectively, with a normalized spectrum brightness of 4.9/s/MHz/mW. Due to the ability of being independently locked to two different wavelenghts, the conjoined double-cavity is universally suitable for hybrid quantum network consisting of various quantum nodes.
	\end{abstract}
	

	\maketitle

The realization of quantum network composed with quantum nodes and quantum channels is of great importance to the distributed quantum computation and quantum communication. The quantum node works for operating and storing quantum state and the quantum channel works for distributing quantum information \cite{1}. Many experimental progresses have been achieved in this aspect with different physical systems, such as atomic ensembles, single atoms and trapped ions \cite{2, 3, 4}. To be compared, hybrid quantum network consisting of various physical systems \cite{5,6} can combine the different advantages of diversities of physical systems. However, the central frequencies of photons for the interfaces linking the nodes built with different physical systems are usually not identical. To solve this problem, the available proposals are coherent quantum frequency conversion \cite{7}, tailoring the frequency of one kind of node to be same with the other one \cite{6} or interconnecting the nodes with nondegenerate photon pairs \cite{8}.

As an advantage of the hybrid quantum network, different physical systems can be chosen as computing or storing nodes respectively. For the computing nodes, trapped ions can be a suitable candidate, because it is one of the most promising physical systems to realize quantum computation and simulation \cite{9, 10}. The creation of Greenberger-Horne-Zeilinger states in ion trap has been up to 14 qubits \cite{11}, and two-dimensional Ising interactions simulated with hundreds of spins \cite{12} has been reported in experiment. To build a quantum network based on trap ions, it should have the ability to interconvert stationary and flying qubits \cite{13}. The emitted photon-ion entanglement and photon-mediated entanglement have been demonstrated in previous experiments \cite{14, 15}. Recently, there have been experimental results showing that single photons can be absorbed by the trapped single ion \cite{6, 16, 17}. However, to the best of our knowledge, the customized narrow-band photon pair source suitable for trapped ions has not been reported yet.
For the storing nodes, a promising candidate is the rare-earth ion-doped crystal system, because it has already shown some excellent merits as a quantum memory \cite{18, 19}. There have been a lot of major advances in rare earth ion-doped crystal based quantum memory, including demonstration of stopped light and image storage up to the regime of one minute \cite{20}, and up to 0.999 process fidelity for the storage and retrieval process of single-photon-level coherent pulse \cite{21}. Moreover, recent progresses include the storage of photonic high-dimensional OAM Entanglement \cite{22}, storage of telecom wavelength time-bin entanglement \cite{23}, quantum teleportation of the polarization state of a telecom-wavelength photon onto the state of a solid-state quantum memory \cite{24}, and the preservation of quantum coherence on an hour-long timescale \cite{25}. In view of these, the realization of photonic channels linking the trapped ions and the rare earth ion-doped crystals should be useful for the hybrid quantum network.
\begin{figure*}[tb]
   \centering
        \includegraphics[width=0.7\textwidth]{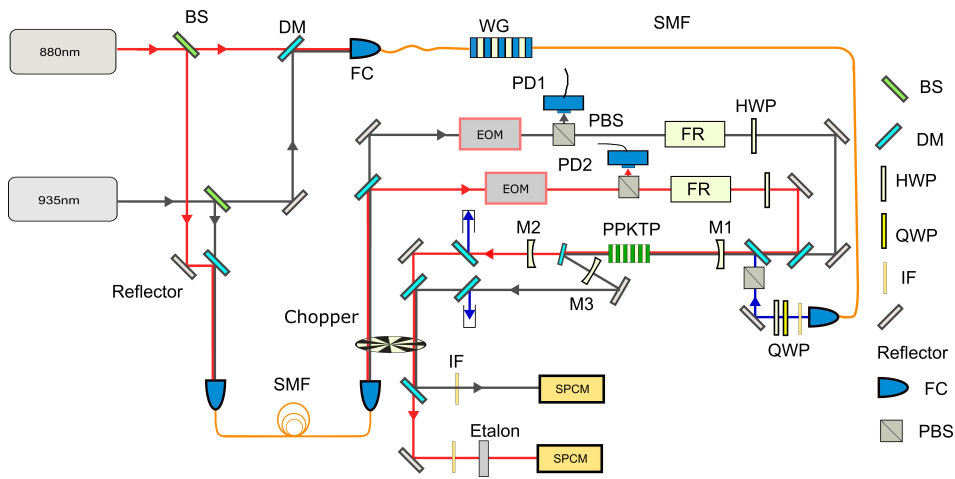}
\caption{Experimental setup for narrow-band photon pair source. BS, beam splitter; HWP, half-wave plate; QWP, quarter-wave plate; PBS, polarization beam splitter; WG, waveguide; EOM, electro-optic modulator; IF, interference filter; FR, farady rotator; FC, fiber coupler; PD1, PD2, photodiodes; SMF, single-mode fiber; SPCM, single photon counting module. The locking beams are frequency-modulated at 50 MHz by EOMs to generate the sidebands for the cavity locking. The combination of HWP and FR is to get the reflected laser from the cavity, and the PDs detect the optical signal. The etalon is used to filter out the multimode components.}
\label{1}
\end{figure*}

Compared to generated photon pairs from cold atomic system \cite{17, 26}, cavity-enhanced spontaneous parametric down-conversion (SPDC) process is much more flexible and less complicated for providing photonic channels. Passive filtering with optical etalon is also a direct way to get band-matched narrow-band photon pairs from single-pass SPDC source \cite{16}, but it will decrease the photon counting rate extremely. During the process of cavity-enhanced SPDC, the down conversion photons with the frequency and spatial mode matched to the cavity mode are greatly enhanced, otherwise greatly suppressed \cite{27}, resulting in generation of photon pairs  with flexible bandwidth and perfect spatial mode. Since the first cavity-enhanced SPDC was demonstrated \cite{27}, lots of experiments have been performed for different purposes \cite{28, 29, 30, 31, 32, 33}. Recently, nondegenerate narrow-band photon pairs are generated with different methods \cite{33, 34, 35}. However, linking two kinds of physical systems requires excellent match of the central frequencies and the bandwidths for each node at the same time. In experiment \cite{33}, the frequency of idler photons can only be some certain values determined by the frequency of the signal photons because of the locking system. In experiments \cite{34, 35}, the frequencies of two photons are changing together with the tuning of temperature. So the requirement can not be fulfilled in above experiments.

In this letter, we demonstrate a new way to generate nondegenerate narrow-band photon pairs with a conjoined double-cavity (CDC) structure, and the central frequency and bandwidth are matched well to the trapped ion $\textrm{Yb}^+$ \cite{6} and the $\textrm{Nd}^{3+}$-doped solid-state quantum memory \cite{21}. The good match of central frequency is perfectly guaranteed by the fact that the two cavities in the CDC structure are independently locked on the resonance absorption lines of two different kinds of nodes, and the bandwidth match is ensured by modified design of two individual cavities. The measured bandwidths of the generated photon pairs is 4 MHz at 935 nm and 5 MHz at 880 nm. To further take advantage of long coherence time of narrow-band photons, single-mode output of photons is realized by the temperature-controlled etalon filter.

The experimental setup of our narrow-band photon pair source is shown in Fig. 1. Two semiconductor ECDL lasers with target wavelengths at 880 nm (Moglabs) and 935 nm (Toptica DL pro) are tuned to the working wavelengths of quantum memory and trapped ion. Two lasers are locked to the same ultrastable Fabry-P\'{e}rot (FP) cavity with 1 MHz bandwidth and 1500 finesses (Stable Laser Systems) with Pound-Drever-Hall scheme, which is not shown here \cite{37}. The 453 nm pump laser is provided by the sum frequency generation (SFG) of the two 880 nm and 935 nm lasers for the next step SPDC process. The SFG process mainly consists of a 1 cm long periodically poled lithium niobate (PPLN) waveguide (HCPhotonics Corp). In the cavity-enhanced SPDC process, a 2 cm long periodically poled KTiOPO4 (PPKTP) crystal is used as the nonlinear crystal. The type-II quasi-phase matching is fabricated for the polarization of 453 nm pump laser being paralleled with the polarization of 935 nm laser. The calculated theoretical bandwidth of the phase-matching is 120 GHz. As shown in Fig. 1, a customized dichroic mirror (DM) is inserted in the optical cavity, which has a high transmissivity at the wavelength of 880 nm ($T\geqslant99.5\%$ for s/p polarization) and a high reflectivity at the wavelength of 935 nm ($R\geqslant99.99\%$ for s/p polarization). The incident angle of the DM is about 10 degree, resulted from the harsh demand for the good performance of two wavelengths at the same time. The cavity mirror M1 is high-reflection-coated to work as an input coupler ($R\geqslant99.8\%$ for 880 nm and 935 nm), and the cavity mirrors M2, M3 work as the output couplers for 880 nm and 935 nm respectively with a PZT attached on each of them ($R=97\%$). Thus the CDC structure is constructed with the help of the inserted dichroic mirror and the input coupler M1. The curvature of all the cavity mirrors is 10 cm, and the FSRs of the two cavities are both about 800 MHz. The nonlinear PPKTP crystal is also shared by the two cavities. This design provides a flexible method to overcome the critical difficulty to match the central frequencies and bandwidths of the signal and idler photons to different quantum systems. All of the optical elements except inserted DM is antireflectively coated at 453 nm such that the pump laser passes the nonlinear PPKTP crystal only once.

To realize the cavity-enhanced SPDC process, the two cavities in our CDC structure are locked to two reference laser beams independently. Owing to the same traveling direction of the locking beam and SPDC photons, the single photon counting module (SPCM) is under the threat of being excessively illuminated. A widely used method is intermittently locking the cavities with the help of a mechanical chopper \cite{31, 33}. However, in nondegenerate case \cite{33}, two mechanical choppers are usually needed to block two different locking beams and the critical phase lock should be achieved between two choppers, due to the inevitable difference frequency generation process in the cavity. In our design, as shown in Fig. 1, two locking beams are combined together, so are the SPDC photon pairs, to pass through the chopper, then separated by the dichroic mirror. These will ensure that the SPCM is definitely safe even if the chopper is broken down or the electricity shuts down. Besides, thanks to the CDC design, no double resonances in one cavity is needed as before \cite{31}, which will decrease the demand for the accuracy of the temperature control of PPKTP crystal. All the above improvements make our setup rather robust against the exotic environmental disturbances.

It has been shown that the SPDC photon pairs emitted from the cavity are often multimode output \cite{27}, which will decrease the coherence and limit the use of the narrow band photon pairs in experiments. So we put an etalon filter after the output coupler of the 880 nm cavity (shown in Fig. 1) to eliminate the extra multimode components \cite{31}. The etalon (central wavelength 880 nm, bandwidth 120 MHz, FSR 8.4 GHz) is placed in a temperature-controlled oven with an accuracy of 10 mK to achieve a good filtering effect.
\begin{figure}[tb]
    \centering
        \includegraphics[width=0.48\textwidth]{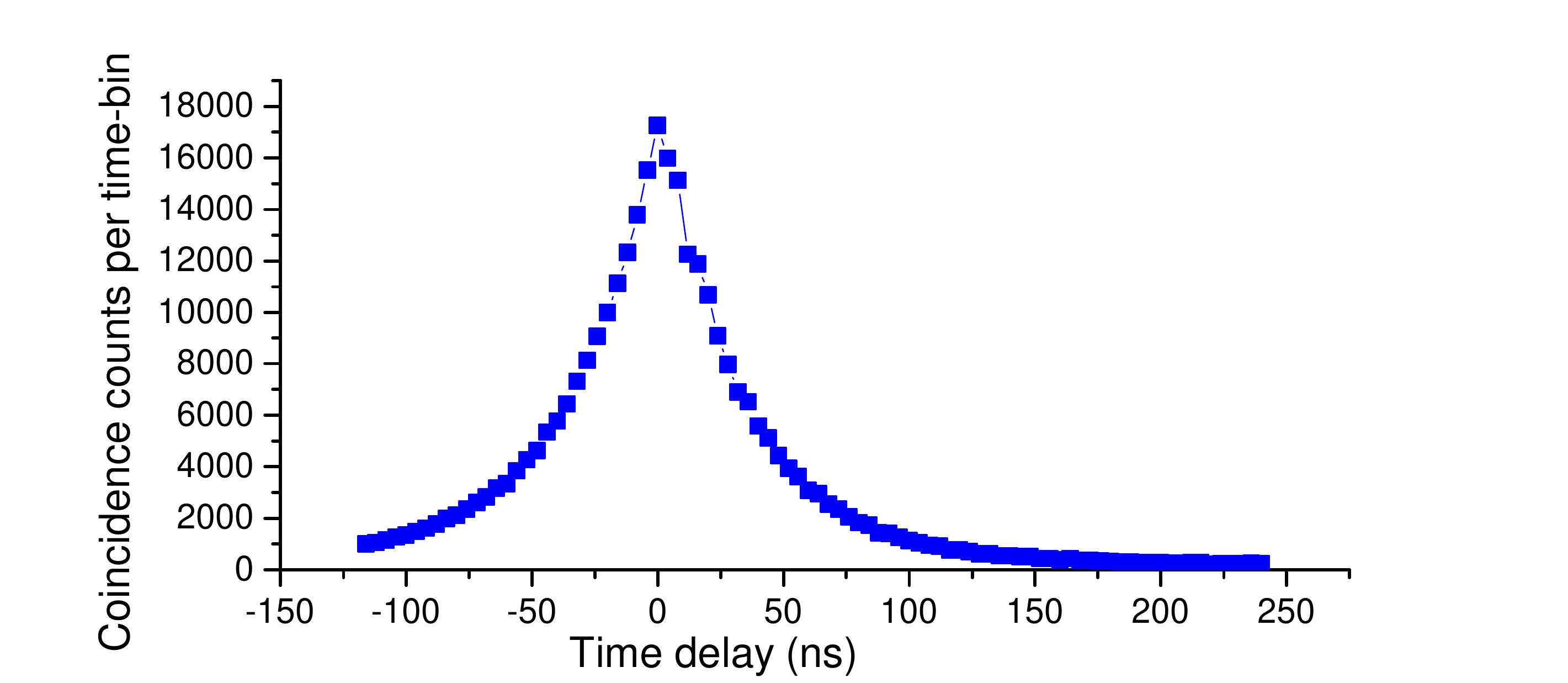}
\caption{The correlation function $G_{S,I}^{(2)}(\tau)$ is measured at 0.8 mW pump power. The time-bin size here is 4 ns and the integration time is 1200 s.
}
\label{fig2}
\end{figure}

To learn more about the generated photon pairs, the time-correlated measurements are conducted in our experiments. The specific method is to measure the second-order cross-correlation function, $G_{S,I}^{(2)}(\tau)$, between photon pairs  \cite{32, 33}. The signal and idler photons are separated by a dichroic mirror and fiber-coupled into the SPCMs for detection after necessary optical filtering. The time distribution of photon pairs arriving at the detectors is recorded by the time-to-digital-converter (Picoquant 400). The measured results of $G_{S,I}^{(2)}(\tau)$ are plotted in Fig. 2. The correlation function is fitted separately on the two sides of curve referred to the fitting function $e^{-2\pi\Delta\upsilon\tau}$, resulting in the signal and idler photons bandwidths being 4 MHz at 935 nm and 5 MHz at 880 nm. The small difference is mainly caused by the different losses of the inserted dichroic mirror at the two wavelengths.

For the purpose of quantum information processing, the high value of normalized cross-correlation function at zero delay, $g_{S,I}^{(2)}(0)$, is very important and necessary. The detail form of normalized cross-correlation function can be expressed as:
\begin{equation}
  g_{S,I}^{(2)}(\tau)=\frac{\langle E_{S}^{\dag}(t)E_{I}^{\dag}(t+\tau)E_{I}(t+\tau)E_{S}(t)\rangle}{\langle E_{I}^{\dag}(t+\tau)(t)E_{I}(t+\tau)\rangle\langle E_{S}^{\dagger}(t)E_{S}(t)\rangle}
\end{equation}
where $E_{S,I}(\tau)$ indicates the operators of signal and idler fields. It is calculated using the correlation function as the peak value divided by the average count value in the accidental region (in the range from 200 ns to 250 ns in Fig. 2 \cite{33}. The measured $g_{S,I}^{(2)}(0)$ varying with different input pump power is plotted in Fig. 3, showing a well fitted inverse proportion relation \cite{35}. The $g_{S,I}^{(2)}(0)$ is about 88 when the input pump power is 1 mW, much higher than  the classical threshold of 2 for two-mode squeezed states. Decreasing the pump power to less than 100 $\mu$W, the $g_{S,I}^{(2)}(0)$ gets a high value of about 800. All these values are calculated without subtracting any background counts. The quite high values of $g_{S,I}^{(2)}(0)$ indicate the good quality of the quantum source and reliable use in hybrid quantum network. The coincidence rate varing with different pump power is also shown in Fig. 3.

The time-resolved measurement $G_{S,I}^{(2)}(\tau)$ of generated cavity-enhanced photon pairs should be a time-domain comblike  structure of the curve, for the reason of  interference between different frequency modes \cite{28}. The period of the interference fringe is \textrm{$t=1/FSR$}, be equal to the cavity round trip time. In our case, as the SPCMs have a resolution time of 350 ps, and the cavity round trip time of 1.25 ns, we get a unperfect comblike interference fringes, as plotted in Fig. 4. The destructive interference part cannot reach zero, resulted from the small ratio between cavity round trip time and resolution time. The asymmetry of interference fringes can be blamed for the mode difference between the two cavities. Then an etalon is inserted to filter the unwanted multimodes. Based on the same time-resolved measurement, the comblike structure of the curve almost disappears, as in the inset of Fig. 4. This relatively smooth curve indicates a quite large elimination of multimode photon pairs.

\begin{figure}[tb]
    \centering
        \includegraphics[width=0.50\textwidth]{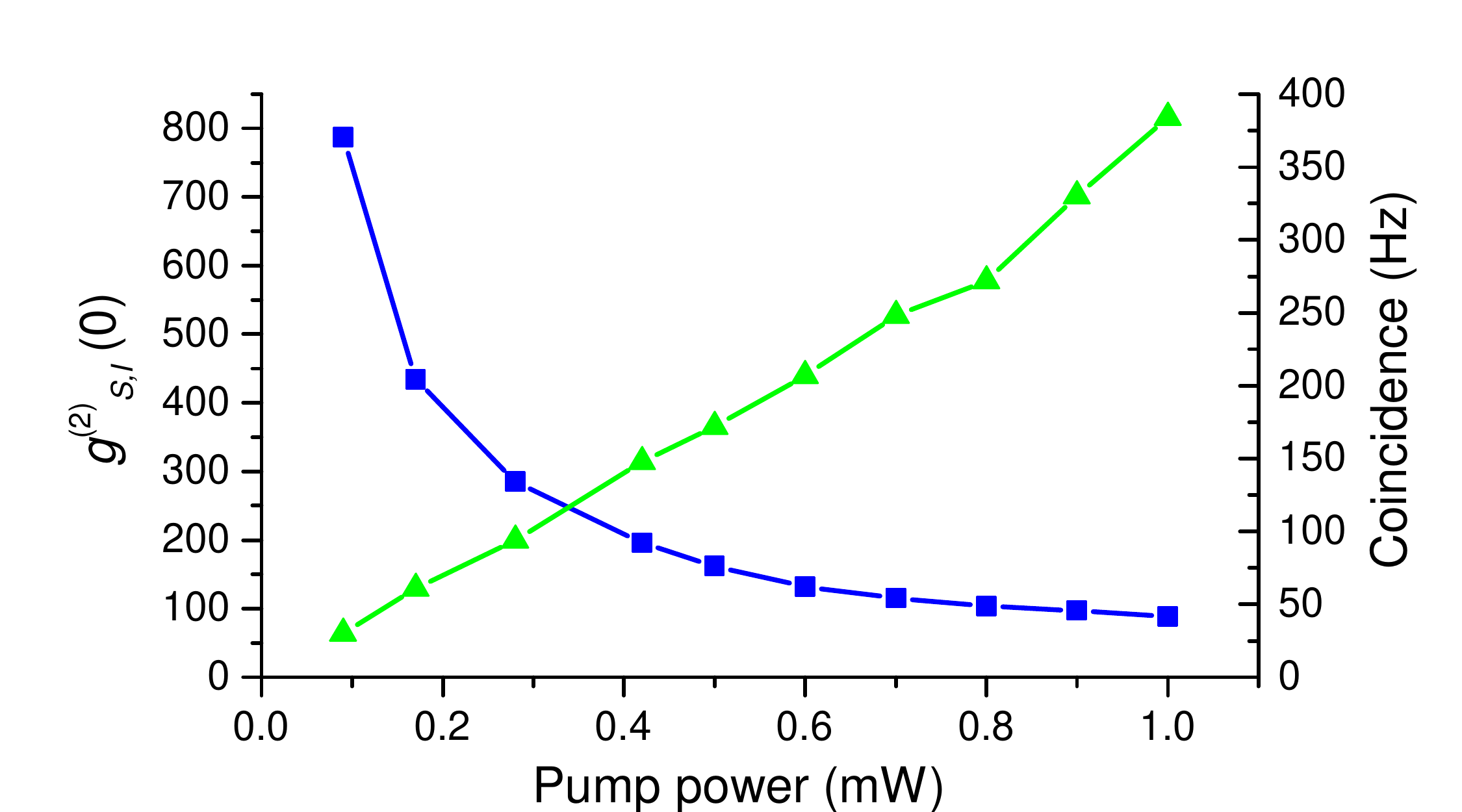}
\caption{The measured $g_{S,I}^{(2)}(0)$ varying with the pump power is plotted at left axis (blue square). The multimode coincidence rate versus the pump power is plotted on right axis (green triangle). The coincidence rate is measured by adding up all the bins of coincidence curve divided by the overall measurement time with the subtraction of background counts \cite{30}.
}
\label{fig3}
\end{figure}

\begin{figure}[tb]
    \centering
        \includegraphics[width=0.50\textwidth]{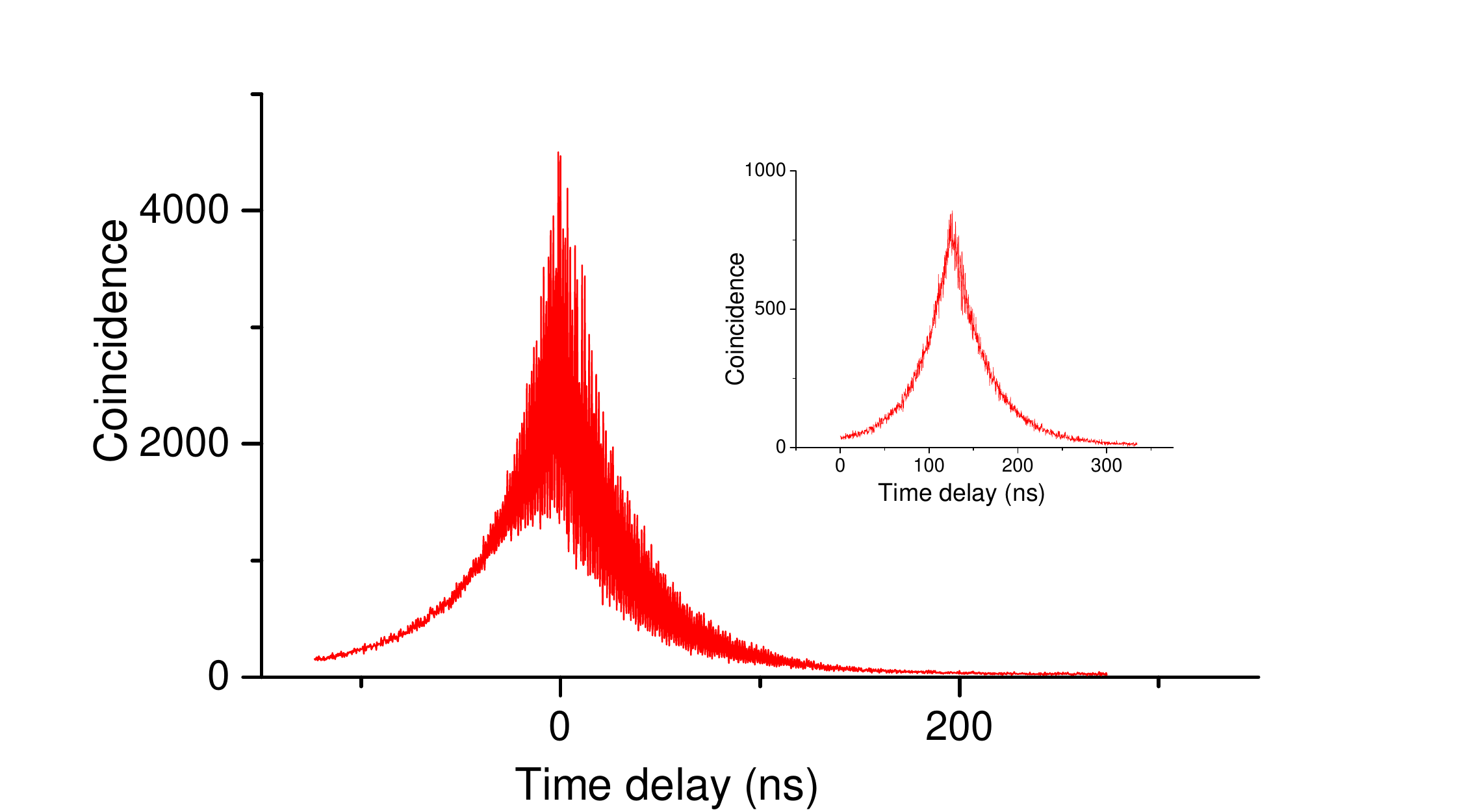}
\caption{The measured time-resolved correlation function $G_{S,I}^{(2)}(\tau)$ with the time-bin size of 256 ps. The imperfect comb-like interference fringes is due to the low ratio between cavity round trip time and resolution time. The inset is measured with the etalon inserted after the 880 nm cavity, showing a good elimination of the multimode components.
}
\label{fig4}
\end{figure}

\begin{figure}[tb]
    \centering
        \includegraphics[width=0.50\textwidth]{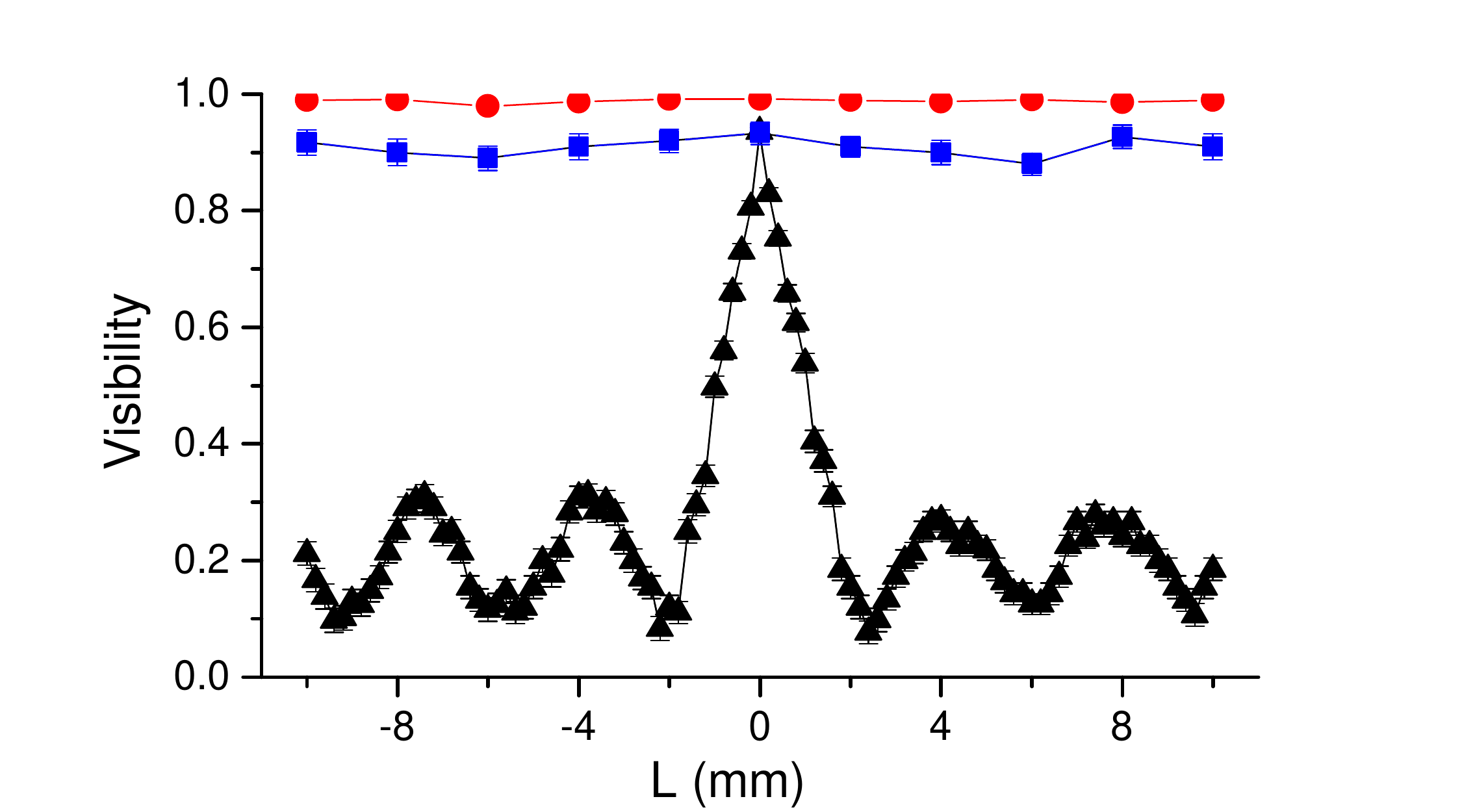}
\caption{Measured results of the visibility of the Michelson interferometer versus its optical path difference in three cases, illustrating the coherence of the input light. High interference visibilities of classical light shows the good quality of the interferometer (red circle). Owing to the existence of multimodes, the unfiltered case shows overall decreasing interference visibilities (black triangle). For filtered case, its interference visibilities are almost as high as classical case, indicating a good filtering of multimodes (blue square).
}
\label{fig5}
\end{figure}

To further ensure the single mode output of generated photon pairs, we use a Michelson interferometer to measure the coherence of the filtered 880 nm photons \cite{33, 36}. The results are shown in Fig. 5. To scan the interference fringes, a reflective mirror in one arm is mounted on a translation stage with a piezoelectric transducer attached for fine tuning. With a 880 nm laser beam fed into the interferometer, a high average interference visibility of about 0.99 is observed, showing the good quality of interferometer. Without the etalon filter, the interference fringes shows decreasing visibilities with a certain period due to the multimode components. While the average interference visibility of the filtered case is $92\%$ in the test region, with a stable trend of high interference visibility when increasing the optical path difference between two arms. The combined measurement of second order cross correlation and first order autocorrelation (coherence) provides a good evidence that our source can provide single-mode output.

After the filtering of multimodes,  the single mode photon pairs generation rate is detected at pump power of $0.9$ mW, which is about $20\textrm{/s}$. Considering the bandwidth mentioned above, we can get the normalized spectrum brightness of about $4.9\textrm{/s/MHz/mW}$ without any modification. To further improve the spectrum brightness, some available methods are to increase the detection system efficiency (including the losses of etalon), the finesse of the cavity and the output proportion of the chopper duty cycle (in our case, $\sim1/3$).

In conclusion, we present a CDC strategy to provide nondegenerate narrow-band photon pairs through cavity-enhanced SPDC process. Based on this strategy, a photon pair source is experimentally generated with narrow bandwidths of 4 MHz at 935 nm and 5 MHz 880 nm, which can be applied to link the ion trap node and solid-state quantum memory node in a hybrid quantum network. Due to the special design of the CDC strategy, the central frequencies and bandwidths of both the signal and idler photons in the pair can be individually set to match these two kinds of nodes very well. This method should be useful for constructing a hybrid quantum network involving nodes of various quantum systems in future.

\vspace{0.1cm}
We thank Z.-Y. Zhou, Y. Li for useful discussion and we also thank Z.-Q. Zhou, G.-Y. Xiang, C.-H. Dong for lending us some experimental devices. This work was supported by the National Natural Science Foundation of China (Nos. 61327901, 61490711, 11274289, 11325419, 61225025, 11474268, 11374288, 11304305, 11404319), the National Basic Research Program of China (No. 2011CB921200), the Strategic Priority Research Program (B) of the Chinese Academy of Sciences (Grant No. XDB01030300), the Fundamental Research Funds for the Central Universities (Nos. WK2470000011, WK2470000018).

\end{document}